\documentclass[twocolumn,aps,prc,superscriptaddress,showpacs]{revtex4}
\usepackage{amsmath,bm}%
\usepackage{graphicx}%

\begin{document}

\draft
\title{ Isoscaling behavior in the Fission Dynamics}


\author{Y. G. Ma} \thanks{Email: ygma@sinap.ac.cn}
\affiliation{Shanghai Institute of Applied Physics, Chinese
Academy of Sciences, P. O. Box 800-204, Shanghai 201800, China}

\author{K. Wang}
\affiliation{Shanghai Institute of Applied Physics, Chinese
Academy of Sciences, P. O. Box 800-204, Shanghai 201800, China}
\affiliation{Graduate School of the Chinese Academy of Sciences,
China}

\author{X. Z. Cai}
\affiliation{Shanghai Institute of Applied Physics, Chinese
Academy of Sciences, P. O. Box 800-204, Shanghai 201800, China}
\author{J. G. Chen}
\author{J. H. Chen}
 \affiliation{Shanghai Institute of Applied
Physics, Chinese Academy of Sciences, P. O. Box 800-204, Shanghai
201800, China} \affiliation{Graduate School of the Chinese Academy
of Sciences, China}
\author{D. Q. Fang}
\affiliation{Shanghai Institute of Applied Physics, Chinese
Academy of Sciences, P. O. Box 800-204, Shanghai 201800, China}
\author{W. Guo}
\author{C. W. Ma}
\author{G. L. Ma}
\affiliation{Shanghai Institute of Applied Physics, Chinese
Academy of Sciences, P. O. Box 800-204, Shanghai 201800, China}
\affiliation{Graduate School of the Chinese Academy of Sciences,
China}
\author{W. Q. Shen}
\affiliation{Shanghai Institute of Applied Physics, Chinese
Academy of Sciences, P. O. Box 800-204, Shanghai 201800, China}
\author{Q. M. Su}
\affiliation{Shanghai Institute of Applied Physics, Chinese
Academy of Sciences, P. O. Box 800-204, Shanghai 201800, China}
\affiliation{Graduate School of the Chinese Academy of Sciences,
China}
 \author{W. D. Tian}
\affiliation{Shanghai Institute of Applied Physics, Chinese
Academy of Sciences, P. O. Box 800-204,    Shanghai 201800, China}
\author{Y. B. Wei}
\author{T. Z. Yan}
\affiliation{Shanghai Institute of Applied Physics, Chinese
Academy of Sciences, P. O. Box 800-204, Shanghai 201800, China}
\affiliation{Graduate School of the Chinese Academy of Sciences,
China}
\author{C. Zhong}
\affiliation{Shanghai Institute of Applied Physics, Chinese
Academy of Sciences, P. O. Box 800-204,   Shanghai 201800, China}
\author{X. F. Zhou}
\affiliation{Shanghai Institute of Applied Physics, Chinese
Academy of Sciences, P. O. Box 800-204,   Shanghai 201800, China}
\affiliation{College of Sciences, Ningbo University, Zhejiang 315211,
China}
\author{J. X. Zuo}
\affiliation{Shanghai Institute of Applied Physics, Chinese
Academy of Sciences, P. O. Box 800-204,  Shanghai 201800, China}
\affiliation{Graduate School of the Chinese Academy of Sciences,
China}

\date{\today}

\begin{abstract}
The fission processes of $^{112}$Sn + $^{112}$Sn and $^{116}$Sn +
$^{116}$Sn are simulated with the combination of the Langevin
equation and the statistical decay model. The masses of two fission
fragments are given by assuming the process of symmetric fission
or asymmetric fission by the Monte Carlo sampling with the
 Gaussian probability distribution.
From the analysis to the isotopic/isotonic ratios of the fission
fragments from both reactions, the isoscaling behavior has been
observed and investigated in details. Isoscaling parameters
$\alpha$ and $\beta$ are extracted as a function of the charge
number and neutron number, respectively,
 in different width of the sampling Gaussian probability distribution.
It seems that $\alpha$ is sensitive to the width of fission
probability distribution of the mass asymmetrical
parameter but $\beta$ is not. Both $\alpha$ and $\beta$
drop with the increasing
of beam energy and the reduced friction parameter.
\end{abstract}



\maketitle

\section{Introduction}

The availability of exotic nuclear beam with extreme
neutron-to-proton ratio provides an opportunity to explore the
collision dynamics of isospin-asymmetric nuclear systems
\cite{RIA}.  To facilitate this kind of study, the suitable
selection of the sensitive experimental observables related to the
isospin degree of freedom is one of key points. One of such
observables is the isotopic/isobaric ratio \cite{Wada,Sherry},
which has been used to probe the isospin equilibration at medium
energies before. Recently, this kind of ratios have been
systematically revisited for two different reactions with the same
charge number and the similar temperature and a so-called
isoscaling law has been observed  experimentally \cite{TsangPRL,
Tsang2,Tsang3}. Isoscaling means that the ratio of isotope yields
from two different reactions, 1 and 2, $R_{\rm 21}(N,Z) =
Y_2(N,Z)/Y_1(N,Z)$, is found to exhibit an exponential
relationship as a function of the neutron number $N$ and proton
number $Z$ \cite{TsangPRL}
\begin{equation}
 R_{\rm 21}(N,Z) = \frac{Y_2(N,Z)}{Y_1(N,Z)} = C exp(\alpha N + \beta
 Z),
\end{equation}
where $C$, $\alpha$ and $\beta$ are three parameters. In
grand-canonical limit, $\alpha = \Delta\mu_{n}/T$ and $\beta =
\Delta\mu_{z}/T$ where $\Delta\mu_{n}$ and $\Delta\mu_{z}$ are the
differences between the neutron and proton chemical potentials for
two reactions, respectively. This behavior is attributed to the
difference of two reaction systems with different isospin
asymmetry. It is potential to probe the isospin dependent nuclear
equation of state by the studies of isoscaling \cite{Ma_review}.
So far, the isoscaling behavior has been  experimentally explored
in various reaction mechanisms, ranging from the evaporation
\cite{TsangPRL}, fission \cite{Friedman,Veselsky2} and deep
inelastic reaction at low energies to the projectile fragmentation
\cite{TAMU1,Veselsky} and multi-fragmentation at intermediate
energy \cite{TsangPRL,LiuTX,Geraci}. While, the isoscaling
phenomenon has been extensively examined in different theoretical
frameworks, ranging from dynamical transport models, such as
Blotzmann-Uehling-Uhlenbeck model \cite{LiuTX}, Quantum Molecular
Dynamics model \cite{Tian}
 and Anti-symmetrical Molecular Dynamics model
\cite{Ono}, to the statistical models, such as the expansion emission
source model, the statistical multi-fragmentation model and the
lattice gas model \cite{Tsang2,Tsang3,Botvina,Souza,Ma}.
In this work, we will focus on the detailed simulation studies on
the isoscaling behavior of the fission fragments.
A brief report has been published recently \cite{Wang}.

In this work, we present an analysis for the fragments from the
fission which was simulated by the Langevin equation. The isotopic
or isotonic ratios of the different fragment yields from
$^{116}$Sn + $^{116}$Sn and $^{112}$Sn + $^{112}$Sn system are
presented and the features of isoscaling behavior in fission
dynamics are investigated.

The paper is organized by the following structure. In Sec. II, a
brief description of the Langevin model is given and the partition
of masses of two fission fragments is assumed; In Sec. III, the
detailed results for the fission-fragment isotopic and isotonic
distribution are presented and the isoscaling behavior is
explored; Finally we summarize the present work.

\section{Brief Description of the Langevin Model}

The process of fission can be described in terms of collective
motion using the transport theory
\cite{Randrup,CDSM,Feldmeier,HHofmann,Frobrich}. The dynamics of
the collective degrees of freedom is typically described using the
Langevin or Fokker-Planck equation. In this work,  we deal with a
Combined Dynamical and Statistical Model (CDSM) which is a
combination of a dynamical Langevin equation and a statistical
model to describe the fission process of heavy ion reaction
\cite{CDSM}. This model is an overdamped Langevin equation coupled
with a Monte Carlo procedure allowing for the discrete emission of
light particles. It switches over to statistical model when the
dynamical description reaches a quasi-stationary regime. We first
specify the entrance channel through which a compound nucleus is
formed, i.e. the target and projectile is complete fusion. The
fusion process of simulating the fission in each trajectory with
angular momentum $L=\hbar l$ is described by

\begin{equation}
\sigma(l)=\frac{2\pi}{k^{2}}\frac{2l+1}{1+exp[(l-l_{c})/\delta_{l}]}
\end{equation}
where the parameters $l_{c}$ and $\delta l$ is according to an
approximating scaling of Ref.~\cite{Frobrich}. Namely,
\begin{widetext}
\begin{equation}
l_c = \sqrt{A_P \times A_T/A_{\rm CN}} \times
(A_P^{1/3}+A_T^{1/3}) \times (0.33 + 0.205 \times \sqrt{E_{\rm c.
m.}-V_c} )
\end{equation}
\end{widetext}
when $0 < E_{\rm c. m.} - V_c < 120 $ MeV; and when $E_{\rm c.
m.}-V_c
> 120$ MeV the term in the last bracket is put equal to 2.5. In
the above equation, $A_{T}$ and $A_{P}$ represents the mass of
target and projectile, respectively, and $A_{\rm CN}$ is the mass
of compound nucleus. For the barrier $V_c$ the following ansatz is
used:
\begin{equation}
V_c = \frac{5}{3} c_3 \times \frac{Z_PZ_T}{A_P^{1/3}+A_T^{1/3}+1.6},
\end{equation}
with $c_3$ = 0.7053 MeV. The diffuseness $\delta l$ is found to scale as
    \begin{widetext}
\[\delta l = \left.\{ \begin{array}{ll}(A_P A_T)^{3/2} \times 10^{-5} \times [1.5 + 0.02 \times (E_{\rm c. m.}-V_c-10)]
& \mbox{for $E_{\rm c. m.} > V_c + 10$}\\(A_P A_T)^{3/2} \times
10^{-5} \times [1.5 - 0.04 \times (E_{\rm c. m.}-V_c-10)] &
\mbox{for $E_{\rm c. m.} < V_c + 10$}\end {array} \right. \]
\end{widetext}

Trajectory with the particular angular momentum $L$ is started at
the ground state position $q_{\rm gs}$ of the entropy $S(q_{\rm
gs},E^{*}_{\rm tot},A,Z,L)$, $q$ is half of the distance between
the centers of masses of the future fission fragments. In this
work the total initial excitation energy $E^{*}_{\rm tot}$ is
given by $E^{*}_{\rm tot} = E_{\rm beam}A_{T}/(A_{T}+A_{P})+ Q$
where $Q$ is the fusion $Q$-value calculated by $Q =
M_{T}+M_{P}-M^{\rm LD}_{\rm CN}$. $M_{T}$ and $M_{P}$ is the mass
of projectile and target come from experimental data,
respectively. If it is unavailable, it is calculated by
macroscopic-microscopic model \cite{macr-micr}. $M^{\rm LD}_{\rm
CN}$ is the mass of the compound nucleus which is calculated from
the liquid-drop model.

The dynamical part of CDSM model is described by the Langevin
equation which is driven  by the free energy $F$. $F$ is related
to the level density parameter $a(q)$ \cite{Ignatyuk}
\begin{equation}
F(q,T) = V(q) - a(q)T^{2}
\end{equation}
in the Fermi gas model, where $V(q)$ is the fission potential and $T$ is the
nuclear temperature.

The overdamped  Langevin equation reads
\begin{eqnarray}
 \frac{dq}{dt} = -\frac{1}{M\beta_0(q)}(\frac{\partial F(q,T)_{T}}{\partial
 q})+\sqrt{D(q)}\Gamma(t),
\end{eqnarray}
where $q$ is the dimensionless fission coordinate defined as above.
 $\beta_0(q)$ is the reduced
friction parameter which is the only parameter of this model.
 The fluctuation strength coefficient $D(q)$ can be expressed
 according to the fluctuation-dissipation theorem:
 \begin{equation}
 D(q) = \frac{T}{M\beta_0 (q)} ,
 \end{equation}
 where    $M$ is the inertia
   parameter which drops out of the overdamped equation.
$\Gamma(t)$ is a time-dependent stochastic
variable with Gaussian distribution. Its average and  correlation
function is written as
\begin{eqnarray}
 <\Gamma(t)> = 0,\nonumber\\
 <\Gamma(t)\Gamma(t')> = 2\delta_{\varepsilon}(t-t').
\end{eqnarray}

The potential energy $V(A,Z,L,q)$ is obtained from the
finite-range liquid drop model \cite{liquid}
\begin{eqnarray}
V(A,Z,L,q) = a_{2}[1-k(\frac{N-Z}{A})^{2}]A^{2/3}[B_{s}(q)-1]\nonumber\\
 +
 c_{3}\frac{Z^{2}}{A^{1/3}}[B_{c}(q)-1]+c_{r}L^{2}A^{-5/3}B_{r}(q),
 \label{Eq_potential}
\end{eqnarray}
where $B_{s}(q)$, $B_{c}(q)$ and $B_{r}(q)$ means surface, Coulomb
and rotational energy terms, respectively, which depends on the
deformation coordinate $q$. $a_{2}$, $c_{3}$, $k$ and $c_{r}$ are
parameters not related to $q$. In our calculation we take them
according to Ref.~\cite{Frobrich}.

\begin{eqnarray}
a_{2} = 17.9439 {\rm MeV}, ~~~c_{3} = 0.7053 {\rm MeV},\nonumber\\
k = 1.7826,~~~~~~~~~~~~~~~c_{r} = 34.50 {\rm MeV}.\nonumber
\end{eqnarray}

We use $c$ and $h$ \cite{ch} to describe the shape of nucleus,
\begin{equation}
\rho^{2}(z) =
(1-\frac{z^{2}}{c_{0}^{2}})((\frac{1}{c^{3}}-\frac{b_{0}}{5})c_{0}^{2}+
B_ {\rm sh}(c,h)z^{2}),
\end{equation}
where
\begin{equation}
{c_{0}} = c R, ~~~~R = 1.16A^{1/3}.
\end{equation}
The nuclear shape function $B_ {\rm sh}(c,h)$ and the collective
fission coordinate $q(c,h)$ of mass number $A$ is expressed as
\begin{eqnarray}
 B_ {\rm sh}(c,h) = 2h + \frac{c-1}{2},\nonumber\\
 q(c,h) = \frac{3}{8}c(1 + \frac{2}{15}B_ {\rm sh}(c,h)c^{3}).
\end{eqnarray}

The fission process of the Langevin equation is propagated using an
interpretation of Smoluchowski \cite{interpretation}($\lambda = 1$
in the following equation) which is consistent with the kinetic
form which reads

\begin{widetext}
\begin{equation}
{q_{n+1}}={q_{n}} +
[\frac{T(q)}{\beta_0(q)M}\frac{dS(q)}{dq}]_{n}\tau +    \lambda
[\frac{d}{dq}(\frac{T(q)}{\beta_0(q)M})]_{n}\tau +
\sqrt{[\frac{T(q)}{\beta_0(q)M}]_{n}\tau}w_{n}
\end{equation}
\end{widetext}

Here $\tau$ is the time step of the Langevin equation, $w_{n}$ is
a Gaussian distributed random number with variance 2.
$S(q)=2\sqrt{a(q,A)[E_{\rm tot}-V(q,A,Z,l)]}$ is the entropy.  The
parameter $\lambda$ allows us to distinguish between the different
possibilities to discretize the Langevin equation. It is called
interpretation in the literature. In the analysis of the
experiments on fission of hot nuclei discussed in the review
\cite{rev1,Frobrich} and in the papers quoted there, the
It\^o-interpretation ($\lambda$ = 0) \cite{Risken} has been used
exclusively. Also there are other interpretations, namely that of
Stratonovich \cite{Stra} ($\lambda$ = 1/2), or an interpretation
which is consistent with the kinetic form of the Smoluchowski
equation of \cite{interpretation} ($\lambda$ = 1). In this work,
we take $\lambda$ = 1.

In our calculation we adopt one-body dissipation (OBD) friction
form factor $\beta_{\rm OBD}$ \cite{obd} as $\beta_0(q)$ which is
calculated with one-body dissipation with a reduction of wall term
except the special case which we claim. Here we use an analytical
fit formula which was developed in Ref. \cite{fit}, i.e.
\[\beta_{\rm OBD}(q)=\left.\{ \begin{array}{ll}15/q^{0.43} + 1 -
10.5q^{0.9} + q^{2} & \mbox{if $q>0.38$}\\32-32.21q & \mbox{if
$q<0.38$}\end {array} \right. \]

In the dynamical part of the model the emission of light particles
(n, p, d, $\alpha$) and giant dipole $\gamma$ are calculated at
each Langevin time step $\tau$, the widths for particle and giant
dipole $\gamma$ decay are given by the parametrization of Blann
\cite{Width} and Lynn \cite{Lynn}, respectively.

\section{Results and Discussions}

\subsection{Isotopic/Isotonic distributions of the fission fragments}

\begin{figure}
\vspace{-0.2truein}
\includegraphics[scale=0.35]{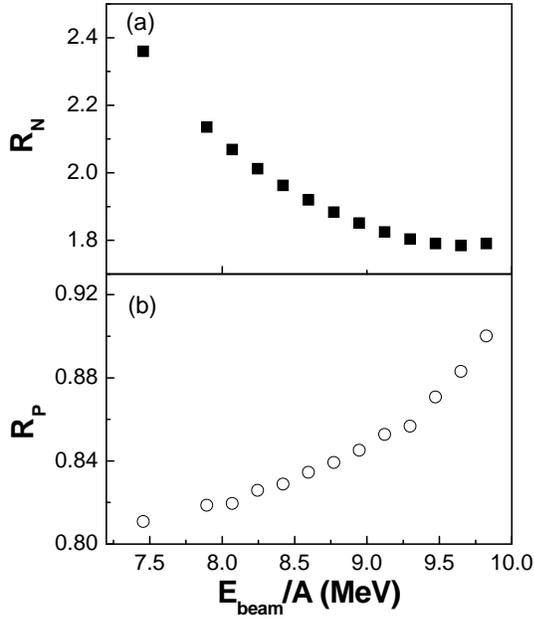}
\vspace{-0.2truein} \caption{\footnotesize The ratios of the
pre-scission neutron number (a) and of the pre-scission proton
number between $^{116}$Sn + $^{116}$Sn and $^{112}$Sn + $^{112}$Sn
as a function of beam energy ($E_{\rm beam}/A$). }
\label{fig_ratio-np}
\end{figure}

In Fig.~\ref{fig_ratio-np} (a) and (b), we demonstrate that the
ratio of the pre-scission neutron number $R_N$ and of the
pre-scission proton number $R_P$ between $^{116}$Sn + $^{116}$Sn
and $^{112}$Sn + $^{112}$Sn, respectively, as a function of beam
energy ($E_{\rm beam}/A$). First, the values of $R_N$ is larger
than 1 while those of   $R_P$ is less than 1, indicating that the
neutron is easier to be emitted for neutron-rich system while the
proton is in contrary trend. Of course, this is a natural result
from the chemical composition of reaction system
\cite{Ma_1999,Ma_1999b}. Second, with the increasing beam energy,
$R_N$ shows a decreasing trend while $R_P$ in reverse way, which
can be interpreted the isospin effect weakens as the beam energy
rises up.

\begin{figure}
\vspace{-0.5truein}
\includegraphics[scale=0.4]{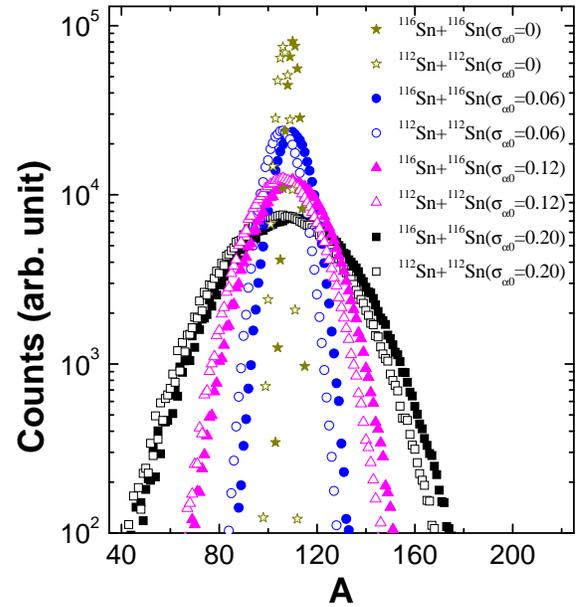}
\vspace{-0.2truein} \caption{\footnotesize (Color online) The
fission-fragment mass distributions produced by the Langevin
simulation for the reactions of $^{112}$Sn + $^{112}$Sn  (open
symbols) and $^{116}$Sn + $^{116}$Sn (filled symbols) at 8.4
MeV/nucleon with the different sampling  width
($\sigma_{\alpha_{0}}$) of Gaussian probability distribution. }
\label{fig_Adist}
\end{figure}

Within the framework of the Langevin simulation we chose 200,000
fission events which happen on dynamic channel (we give up the
events which happen in statistical part of CDSM model) and chose a
Gaussian distributed random number as the mass asymmetry parameter
$\alpha_0 = \frac{A_{1}-A_{2}}{A_{1}+A_{2}}$ when the system
reaches to the scission point. When $\alpha_0 = 0$ it means the
symmetrical fission. It is taken from a Gaussian distributed
random number from -1 to 1 with the mean value of zero. $A_{1}$
and $A_{2}$ is the mass of the two fission fragments,
respectively. In this work we assume the fission fragments have
the same $N/Z$ as the one of the initial system and then $Z_1$ or
$Z_2$ of fission fragments can be deduced from $A_{1}$ or $A_2$.
This assumption is similar to the case of deep inelastic heavy ion
collisions at low energies, where the isospin degree of freedom
has been found to  reach equilibrium first \cite{isospin}.
Fig.~\ref{fig_Adist} shows the mass distribution of the fission
fragments from $^{112}$Sn + $^{112}$Sn and $^{116}$Sn + $^{116}$Sn
reaction systems assuming the different width of the sampling
Gaussian probability for the  mass asymmetrical parameter of
fission fragments ($\sigma_{\alpha_{0}}$). Naturally, the bigger
the $\sigma_{\alpha_{0}}$, the wider the fragment mass
distributions.

\begin{figure}
\vspace{-0.2truein}
\includegraphics[scale=0.4]{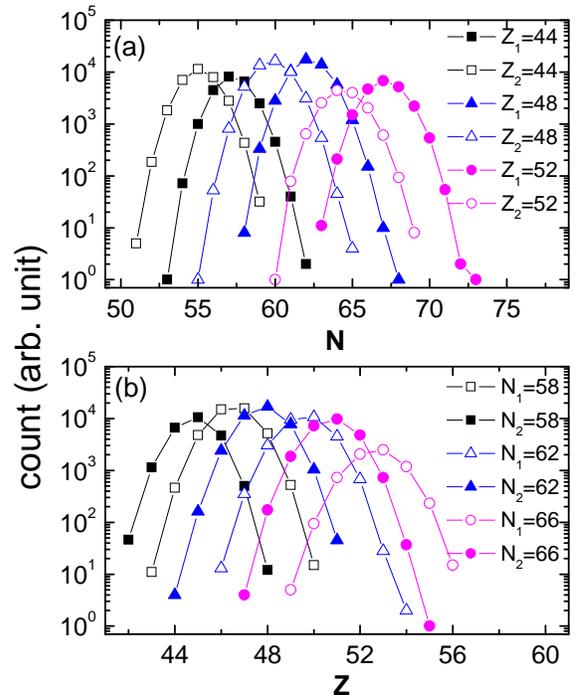}
\vspace{-0.3truein} \caption{\footnotesize (Color online) The
isotopic (a)  and  isotonic (b) distributions of fission-fragments
in some given $Z$ or $N$ (see texts in the inserts) from the
collisions of $^{116}$Sn + $^{116}$Sn (filled symbols) and of
$^{112}$Sn + $^{112}$Sn (open symbols) at 8.4 MeV/nucleon.
$\sigma_{\alpha_{0}}$ = 0.06. Notice that the scale of X-axis is
different. } \label{fig_iso}
\end{figure}

\begin{figure}
\vspace{-0.25truein}
\includegraphics[scale=0.4]{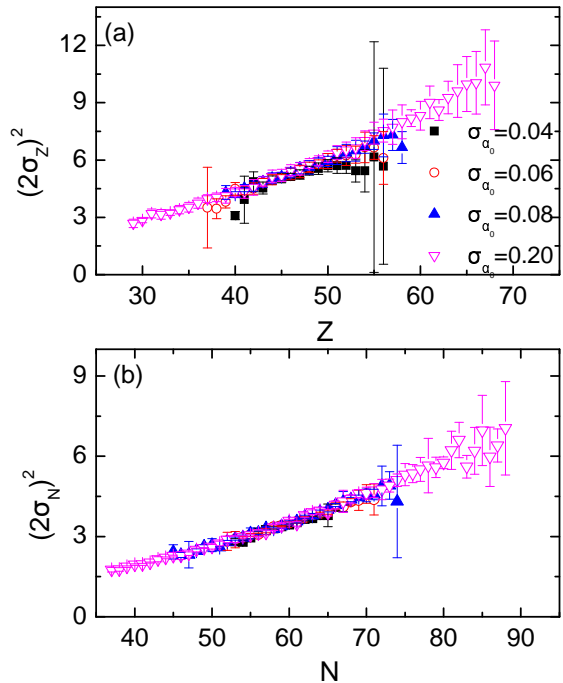}
\vspace{-0.2truein} \caption{\footnotesize (Color online) The
square of full width of the isotopic (a) and isotonic distribution
(b) as a function of $Z$ and $N$, respectively, in different
$\sigma_{\alpha_{0}}$. The width is the average value of the two
systems (they are almost the same for both systems).}
\label{fig_width}
\end{figure}

Samples for the isotopic and isotonic distributions in some given
$Z$ and $N$ are shown in Fig.~\ref{fig_iso}. The square of the
full widths of these distributions shows a systematic increase
with the $Z$ or $N$ as shown in Fig.~\ref{fig_width} and the
absolute value of the differences of the centroid of the
isotonic/isotopic distributions shows an increasing trend too (see
Fig.~\ref{fig_Delta-mun}). Apparently, the widths are not
sensitive to the width of the Gaussian probability for the mass
asymmetrical parameter of fission fragments, but the differences
of the centroid of the isotonic/isotopic distributions shows the
dependence on it.

From a practical point of view, the isoscaling behavior occurs
when two mass distributions for a given $Z$ from two processes
with different isospin are Gaussian distributions with the same
width but different mean mass. In this case, the isotopic
distribution in a given $Z$, namely $Y(N)|_Z$, and isotonic
distribution in a given $N$, namely $Y(Z)|_N$, can be described by
single Gaussian distribution,    respectively, i.e.:
\begin{eqnarray}
Y(N)|_Z \sim exp[-\frac{(N-N_Z)^2}{2\sigma_Z^2}]\nonumber , \\
Y(Z)|_N \sim exp[-\frac{(Z-N_N)^2}{2\sigma_N^2}],
\end{eqnarray}
where $N_Z$ and $N_N$ are the centroid of isotopic and isotonic
distributions,  $\sigma_Z^2$ and $\sigma_N^2$ describe the
variance of distributions for each element of charge $Z$ and
neutron number $N$, respectively. This leads to an exponential
behavior of the ratio $R_{\rm 21}$ if the quadratic term in $N_Z$
or $N_N$ is neglected, it reads
\begin{eqnarray}
ln(R_{\rm 21}(N)|_Z) \sim  \frac{[(N_Z)_2-(N_Z)_1] N}{\sigma_Z^2}\nonumber , \\
ln(R_{\rm 21}(Z)|_N) \sim  \frac{[(N_N)_2-(N_N)_1] Z}{\sigma_N^2}.
\label{appro}
\end{eqnarray}
Note that Eq.(\ref{appro}) requires the values for $\sigma_Z^2$ or
$\sigma_N^2$ to be approximately the same for both reactions,
which is a necessary condition for isoscaling. Indeed, we observed
this case in our simulations for both Sn + Sn collisions. In the
Langevin equation, $\sigma_Z^2$ or $\sigma_N^2$  essentially
depends  on the physical conditions reached, such as the
temperature, the density and the friction parameter etc.
Considering that $R_{\rm 21}(N)|_Z \sim exp(\alpha N)$ or $R_{\rm
21}(Z)|_N \sim exp(\beta Z)$ for a given $Z$ or $N$, we can get
\begin{eqnarray}
\alpha \sim \frac{(N_Z)_2-(N_Z)_1}{\sigma_Z^2}\nonumber , \\
\beta \sim \frac{(N_N)_2-(N_N)_1}{\sigma_N^2}. \label{Eq_sgz}
\end{eqnarray}

Assuming  other ingredients can be neglected, $\sigma_Z^2$ or
$\sigma_N^2$ could be  considered to be proportional to
temperature $T$ of the fission-fragments according to the
fluctuation-dissipation theorem \cite{Feldmeier}, in this
circumstance,
\begin{eqnarray}
\alpha \sim \frac{(N_Z)_2-(N_Z)_1}{T}\nonumber , \\
\beta \sim \frac{(N_N)_2-(N_N)_1}{T} ,
\end{eqnarray}
where $[(N_Z)_2-(N_Z)_1]$ or $[(N_N)_2-(N_N)_1]$ can be understood
as a term of the average difference of the neutron or proton
chemical potential between two reactions.

As we already showed that both $\sigma_Z^2$ and $\sigma_N^2$ rise
with $Z$ and $N$, respectively, and are almost independent of
$\sigma_{\alpha_{0}}$ from Fig.~\ref{fig_width}(a) and (b) in our
calculation. On the other hand, we recognize that  the similar
behavior of the $Z$-dependence of $\sigma_Z$ has been
experimentally observed in the spallation-fission data of
$^{208}$Pb  (1 GeV/nucleon) + $d$ or $p$ etc in Gesellschaft f\"ur
Schwerionenforschung (GSI) \cite{GSI1,GSI2}. According to the
model which is based on the modern version of Abrasion-ablation
model involving the fission nuclei by Benlliure {\it et\ al.}
\cite{GSI3}, the square of the width of symmetric fission fragment
from the macroscopic potential can be expressed by
\begin{equation}
\sigma_{Z}^2 = \frac{1}{2} \frac{\sqrt{E^*_{\rm bf}}}{\sqrt{a}
C_{\rm mac}} = \frac{T_{\rm fis}}{2C_{\rm mac}},
\end{equation}
where ${E^*_{\rm bf}}$ is the excitation energy above the the
fission barrier, $a$ is level energy parameter, $T_{\rm fis}$ is
the temperature of fissioning nuclei and $C_{\rm mac}$ is the
curvature of macroscopic potential energy $V_{\rm mac}$ as a
function of charge asymmetry. In this way, the width of symmetric
fission fragment distribution increases with temperature. This is
also the case in our present model calculation. In the other word,
the temperature of the fission-fragments which mostly originates
from the symmetric fission seems to increases with the charge
number of fragments. Recently, a systematic study on the
experimental data also displays that the variance of the fragment
mass distribution increases with the temperature of the compound
nucleus and the fission-fragments \cite{Saw}.

\begin{figure}
\vspace{-0.25truein}
\includegraphics[scale=0.4]{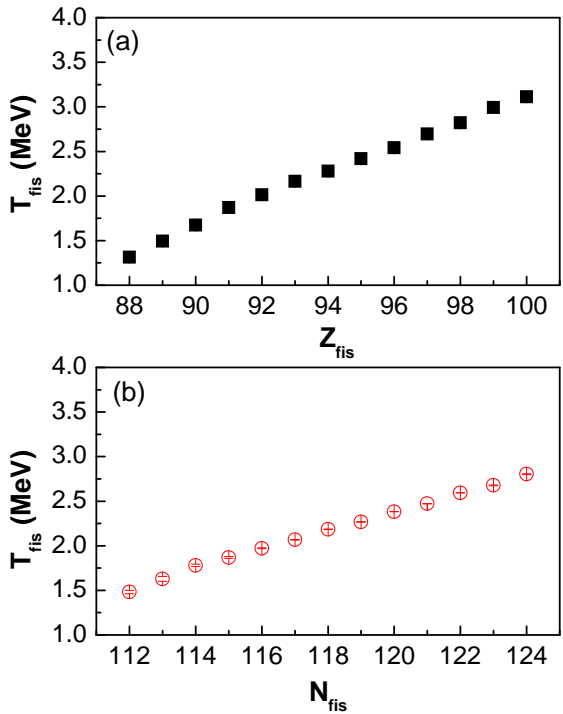}
\vspace{-0.2truein} \caption{\footnotesize (Color online) The
extracted  mean temperature  of the fissioning nuclei of two
reaction systems as a function of their charge number (top panel)
or the neutron number (bottom panel). } \label{Temp_Z}
\end{figure}

\begin{figure}
\vspace{-0.25truein}
\includegraphics[scale=0.4]{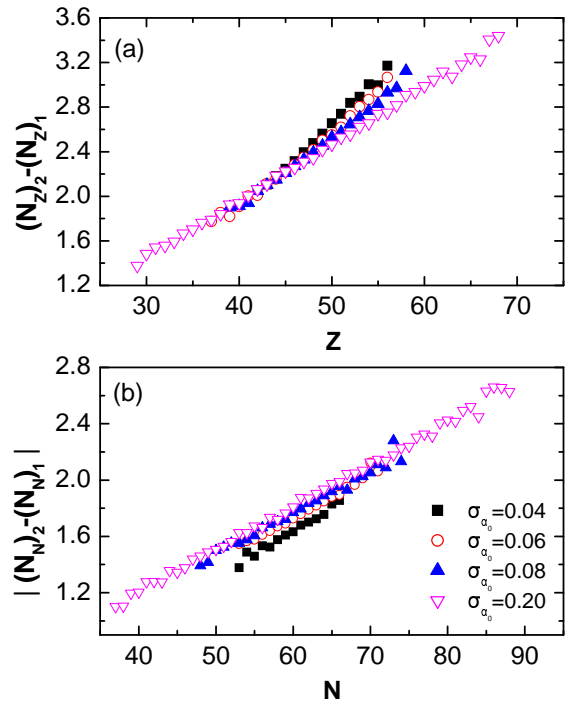}
\vspace{-0.2truein}
 \caption{\footnotesize The absolute value of
  $(N_Z)_2-(N_Z)_1$ (a) and $|(N_N)_2-(N_N)_1|$ (b) as
  a function of $Z$ and $N$ of the fission-fragments, respectively, between $^{116}$Sn + $^{116}$Sn
  and $^{112}$Sn + $^{112}$Sn  with different    $\sigma_{\alpha_{0}}$. }
 \label{fig_Delta-mun}
\end{figure}

To verify the relationship of the temperature and charge number of
fragments as stated above, we extract the temperature of the
fissioning nuclei in the scission point when the system happens on
dynamic channel. Fig.~\ref{Temp_Z}(a) and (b) demonstrate that the
mean temperature of two systems as a function of $Z$ and $N$ for
the fissioning nuclei, respectively. Obviously, the temperature
almost increases linearly with the charge number ($Z_{\rm fis}$)
or neutron number ($N_{\rm fis}$) of the fissioning nuclei. Since
we assume the fission-fragments have the same $N/Z$ as the one of
the fissioning nuclei, hence  the temperature of the
fission-fragments shall increase with their charge number.

\begin{figure}
\vspace{-0.25truein}
\includegraphics[scale=0.4]{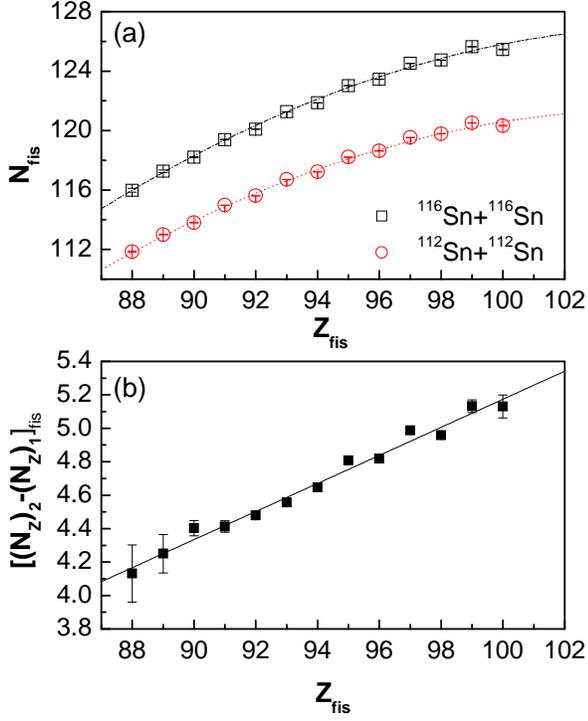}
\vspace{-0.2truein} \caption{\footnotesize (Color online) (a) The
calculated neutron numbers as a function of the charge numbers for
the fissioning nuclei of two reaction systems. (b) The difference
of the neutron number of the fissioning nuclei between two
reaction systems as a function of the charge number for the
fissioning nuclei. The lines represent two-order Polynomial fits
in (a) and a linear fit in (b). } \label{Nz_fis}
\end{figure}

In Eq.(16) $[(N_Z)_2-(N_Z)_1]$ or $[(N_N)_2-(N_N)_1]$  can be
understood as a term of the average difference of the neutron or
proton chemical potential between two reactions.
Fig.~\ref{fig_Delta-mun}(a) and (b) shows $[(N_Z)_2-(N_Z)_1]$ and
the absolute value of $[(N_N)_2-(N_N)_1]$ as a function of $Z$ or
$N$, respectively, in different $\sigma_{\alpha_{0}}$. Apparent
increasing behavior with $Z$ or $N$ has been observed. In order to
understand the increasing behavior of $[(N_Z)_2-(N_Z)_1]$ or
$|[(N_N)_2-(N_N)_1]|$ as a function of the charge number and the
neutron number  of fission-fragments, we investigate the
fissioning nuclei. For an example, Fig.~\ref{Nz_fis}(a) shows the
neutron number versus the charge number of the fissioning nucleus
for both reaction systems just before the fission takes place. The
lines represent the second order polynomial fits to guide the
eyes. From the above points, we can extract the
$[(N_Z)_2-(N_Z)_1]|_{\rm fis}$ as a function of $Z_{\rm fis}$ as
shown in Fig.~\ref{Nz_fis}(b). Since the fission-fragments are
assumed to have the same $N/Z$ as one of the fissioning nuclei,
the fission-fragments shall show the similar increasing behavior
as the charge number rises. In comparison to the insensitivity of
$\sigma_Z$ or $\sigma_N$ to $\sigma_{\alpha_{0}}$,
 $[(N_Z)_2-(N_Z)_1]$ and $[(N_N)_2-(N_N)_1]$ shows somehow
 stronger dependence on   $\sigma_{\alpha_{0}}$, i.e.
$[(N_Z)_2-(N_Z)_1]$ shows larger values in the proton-rich side
while $|[(N_N)_2-(N_N)_1]|$ shows  smaller values in the
neutron-deficit side as $\sigma_{\alpha_{0}}$ becomes smaller in
Fig.~\ref{fig_Delta-mun}. This essentially originates from the
overlap of the different mass partition of two fission-fragments
according to the sampling of Gaussian probability distribution for
the mass asymmetry when the fission takes place. The symmetric
fissions result in the strongest dependence of $[(N_Z)_2-(N_Z)_1]$
or $|[(N_N)_2-(N_N)_1]|$ on the charge number or neutron number.

\subsection{Isoscaling behavior}

\begin{figure}
 \vspace{-0.9truein}
\includegraphics[scale=0.4]{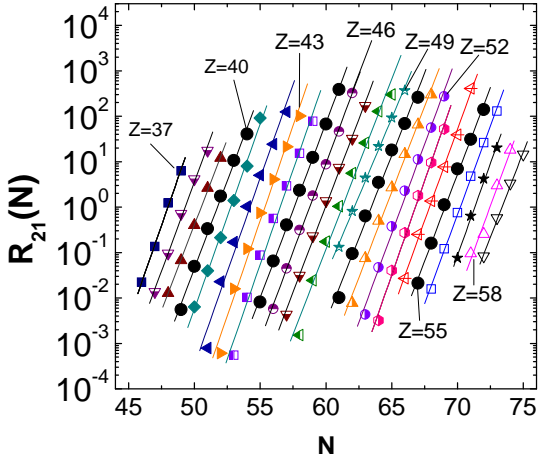}
\vspace{-0.3truein}
 \caption{\footnotesize (Color online) The isotopic yield ratio of the fission fragments
 between $^{116}$Sn + $^{116}$Sn and  $^{112}$Sn + $^{112}$Sn in the
Langevin model with $\sigma_{\alpha_0} = 0.06$ and  $E_{\rm
beam}$/A = 8.4 MeV. Different symbols from left to right represent
the calculated results for the isotopes from $Z$ = 37 to 59. The
lines represent exponential fits to guide the eye.}
\label{fig_isoscaling}
\end{figure}

\begin{figure}
 \vspace{-0.7truein}
\includegraphics[scale=0.4]{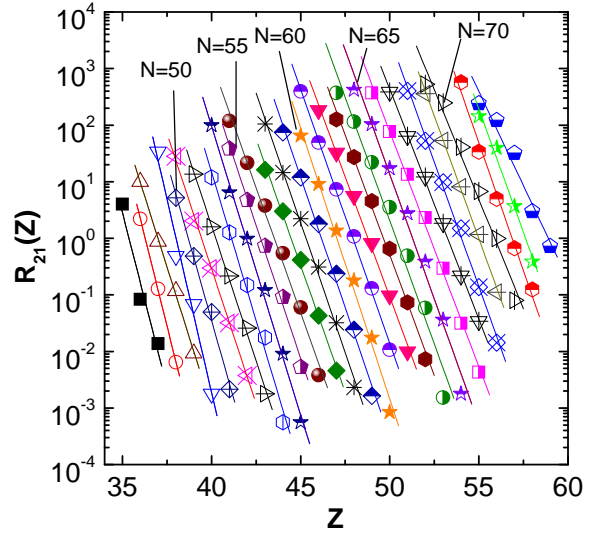}
\vspace{-0.4truein}
 \caption{\footnotesize (Color online) The isotonic yield
 ratio of the fission fragments between  $^{116}$Sn + $^{116}$Sn
 and $^{112}$Sn +
 $^{112}$Sn in the Langevin simulations at $E_{\rm beam}/A$ = 8.4 MeV.
 Different symbols from
 left to right represent the calculated results for the
 isotopes from $N$ = 46 - 73. The width of Gaussian probability is
 assumed 0.06.
 The lines represent exponential fits to guide the eye.
} \label{fig_isobaric}
\end{figure}

Eq.(1) can be written as ${\rm ln}R_{\rm 21} = C_Z + \alpha N$,
where $C_Z = {\rm ln}C + \beta Z$, if we plot $R_{\rm 21}$ as a
function of N, on a natural logarithmic plot, the ratio follows
along a straight line. In Fig.~\ref{fig_isoscaling} this
isoscaling behavior is observed in the Langevin   simulation. Here
each kind of symbol with a line represents a    chain of isotope.
From there, the isoscaling parameter $\alpha$ can be extracted
directly. Similarly, the isoscaling parameter $\beta$ can be
extracted from the isotonic ratio as shown in
Fig.~\ref{fig_isobaric} by
 ${\rm ln} R_{\rm 21} = C_N + \beta Z$, where $C_N = {\rm ln}C + \alpha N$.

\begin{figure}
\vspace{-0.2truein}
\includegraphics[scale=0.4]{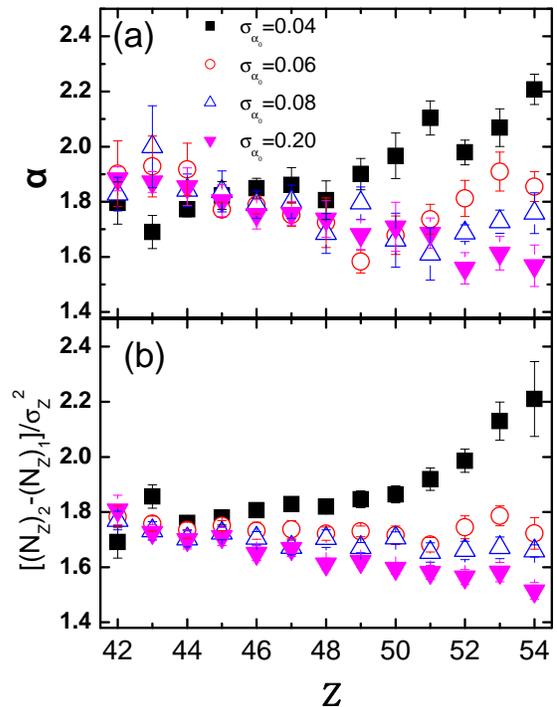}
\vspace{-0.15truein}
 \caption{\footnotesize (Color online) (a) The isoscaling parameter $\alpha$ as
a function of $Z$ in the different Gaussian width ($\sigma_{\alpha_0}$) of
the mass asymmetry parameter $\alpha_0$
for fission fragments; (b) Same as (a) but for
$\frac{[(N_Z)_2-(N_Z)_1]}{\sigma_Z^2}$.} \label{fig_alpha_N_sgmX}
\end{figure}

From Fig.~\ref{fig_isoscaling} and Fig.~\ref{fig_isobaric}, the
relationship between $\alpha$ ($|\beta|$) and the charge number
$Z$ ($N$) of the fission fragments can be deduced. In order to
investigate the effect of the width of Gaussian probability
distribution on the isoscaling parameters, we use the different
widths of the sampling Gaussian distribution for mass asymmetry
parameter $\alpha_{0}$, namely $\sigma_{\alpha_0}$ = 0.04, 0.06,
0.08 and 0.20, with the random number from -1 to 1 and the mean
value of 0. Fig.\ref{fig_alpha_N_sgmX}(a) shows the isoscaling
parameter $\alpha$ as a function of $Z$ with
 different $\sigma_{\alpha_0}$. From this figure, we
know in the low $\sigma_{\alpha_0}$, i.e., when the symmetric
fission is an overwhelming mechanism, $\alpha$ increases with $Z$.
This means that the isospin effect becomes stronger with the
increasing of $Z$. In a recent analyse of fission with a simple
liquid-drop model \cite{Friedman}, a systematic increase of the
isoscaling parameter $\alpha$ with the proton number of the
fragment element has been predicted.  In our simulation, this kind
of increase of $\alpha$ with $Z$ apparently stems from the
dominated symmetric fission mechanism. While, in the another
extreme case from the Fig.~\ref{fig_alpha_N_sgmX}(a), i.e. with
the larger $\sigma_{\alpha_0}$, $\alpha$ shows a contrary trend
with $Z$, i.e., it drops with $Z$. In this case, it seems that
there exists stronger isospin effect for the fragments with lower
$Z$. In a medium case, the rising branch and falling branch
competes with each other, the mediate isoscaling behavior appears
and a minimum of $\alpha$ parameter occurs around the symmetric
fission point. We note that the fission data of $^{238,233}$U
targets induced by 14 MeV neutrons reveal the backbending behavior
of the isoscaling parameter $\alpha$ around the symmetric fission
point \cite{Veselsky2} as stated above. They interpreted that it
originates from the temperature difference of fission fragments
since the isoscaling parameter is typically, within the
grand-canonical approximation, considered inversely proportional
to the temperature ( $\alpha$ = $\Delta \mu_{n}/T$ ) as stated
above. In our case, this kind of backbending of isoscaling
parameter $\alpha$ apparently stems from the moderate width of the
probability distribution of the mass asymmetrical parameter of the
fissioning nucleus as shown in Fig.\ref{fig_alpha_N_sgmX}. or in
the other words, it may stem from a moderate mixture of the
different weights between the symmetric and asymmetric fission
components. Essentially the backbending originates
 from the competition between the term of chemical potential
 and the term of temperature since both terms increase with the
 charge number of fission fragments.

Besides the above direct method to extract isoscaling parameter,
we can also  check the behavior of $\alpha$ in terms of
Eq.~(\ref{Eq_sgz}). Fig.~\ref{fig_alpha_N_sgmX}(b) shows the
$\frac{[(N_Z)_2-(N_Z)_1]}{\sigma_Z ^2}$ as a function of $Z$. With
the increasing of $\sigma_{\alpha_0}$, the $Z$ dependence of
$\frac{[(N_Z)_2-(N_Z)_1]}{\sigma_Z ^2}$ shows from the upswing
trend to downswing trend. A turning point around $Z$ = 51 is also
observered in medium $\sigma_{\alpha_0}$ as
Fig.~\ref{fig_alpha_N_sgmX}(a) shows. From the similarity of  the
behavior shown in Fig.~\ref{fig_alpha_N_sgmX}(a) and (b) as well
as  the approximate equality of the values of $\alpha$ and
$\frac{[(N_Z)_2-(N_Z)_1]}{\sigma_Z ^2}$, we can say that the
Eq.(\ref{Eq_sgz}) works well in the present calculation. In our
case, the turning point of $\alpha$ stems from the competition
between the chemical potential term ($[(N_Z)_2-(N_Z)_1$) and the
temperature term ($\sigma_Z ^2$). In general, the chemical
potential term is more sensitive to the Gaussian width of the mass
asymmetry parameter $\alpha_0$ for fission fragments (see
Fig.~\ref{fig_Delta-mun}). Overall speaking, we find that the
isoscaling parameter $\alpha$ is sensitive to the width of the
probability distribution of mass asymmetrical parameter of the
fission fragments. In the other word, we may say that the
isoscaling parameter is sensitive to asymmetrical extent of both
fission fragments.

\begin{figure}
\vspace{-0.55truein}
\includegraphics[scale=0.42]{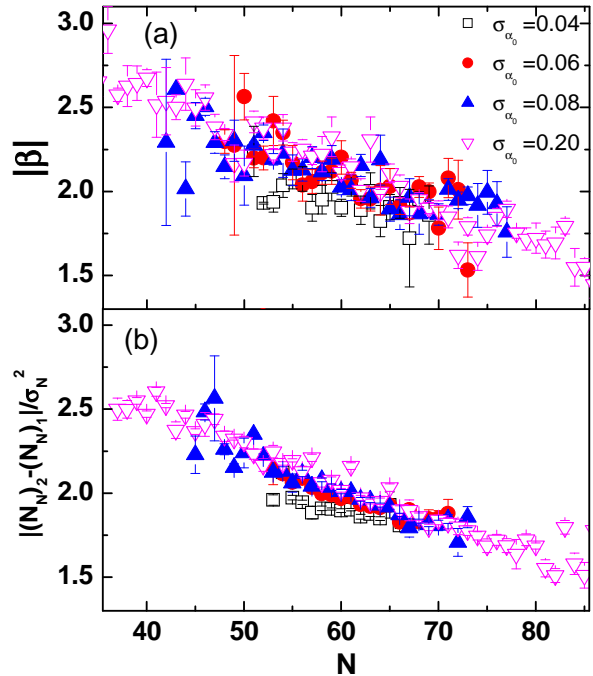}
\vspace{-0.15truein} \caption{\footnotesize (Color online) (a) The
isoscaling parameter $|\beta|$ as a function of $Z$ in the different
 Gaussian width ($\sigma_{\alpha_0}$) of the mass asymmetry parameter
$\alpha_0$ for fission fragments; (b)
Same as (a) but for  $|\frac{(N_N)_2-(N_N)_1}{\sigma_N^2}|$.}
\label{fig_beta_N}

\end{figure}

Similarly, from Figure.~\ref{fig_isobaric}, the relationship between $|\beta|$ and
the neutron number $N$ of the fission fragments can be deduced in
different width $\sigma_{\alpha_0}$. This is shown in Fig.~\ref{fig_beta_N}(a).
Different from the
relationship of $\alpha$ and $Z$, the $|\beta|$ always drops with
the neutron number, regardless of the change of $\sigma_{\alpha_0}$.
The quantitative  and qualitative  similarity of  $\frac{[(N_N)_2-(N_N)_1]}{\sigma_N ^2}$
vs $N$ (Figure.~\ref{fig_beta_N}(b)) has also been observed.
i.e., it always decreases with   $N$ and is insensitive to
$\sigma_{\alpha_0}$.

 However, the obtained isoscaling parameters are actually very large
 in comparison to the usual isoscaling parameter extracted from the data \cite{Veselsky2}.
 The reasons could be the model itself since the model
 is still too simple as well as our assumption of Gaussian
 probability distribution of fission fragments. Also the
 post-fission evaporation component will of course play some
 roles for modification the isoscaling parameters. In the present model calculation, however,
 this influence of post-fission evaporation of
 fission fragment is not included. Those may show larger apparent isoscaling
 parameters in comparison to the data. Of course,  main aim of this
 work is to show the isoscaling behavior of fission fragments and its
 trend with the  charge or neutron number of the fragments by the Langevin dynamics.

 \subsection{The beam energy dependence of the isoscaling parameters}

\begin{figure}
 \vspace{-0.2truein}
 \includegraphics[scale=0.4]{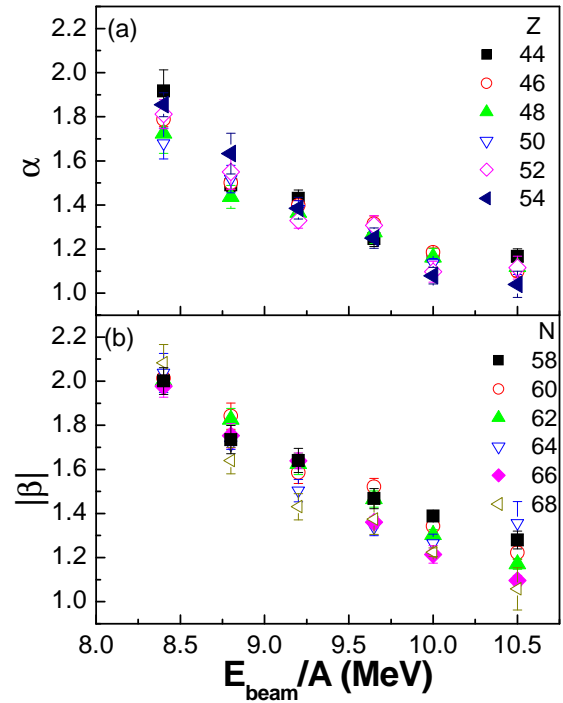}
 \vspace{-.truein} \caption{\footnotesize (Color online) Isoscaling
 parameter $\alpha$ (a) and $\beta$ (b) as a function of beam
 energy for the fragments $Z$ = 44 - 54 and $N$ = 58 - 68,
 respectively.
 The width ($\sigma_{\alpha_0}$) of the Gaussian probability
  is 0.06.
  } \label{fig3}
 \end{figure}

 The simulations are systematically performed in different
 beam energies. The values of $\alpha$ and $\beta$ are extracted as
 a function of beam energy for the fragments $Z$ = 44 - 54 and $N$ =
 58 - 68, respectively, as shown in Fig.\ref{fig3}(a) and (b). It
 shows that both $\alpha$ and $\beta$ decrease as the increasing
 beam energy which means that the isospin effect fades away with the
 increasing of $E_{\rm beam}/A$. This behavior is similar to the case in the
 fragmentation where the
 isoscaling parameter drops with the temperature in the statistical
 models as well as experiments \cite{Tsang3,Ma,Ma_1999b,Ma_CP}.

\subsection{The friction parameter dependence of the isoscaling
parameters}

\begin{figure}
\vspace{-0.1truein}
\includegraphics[scale=0.4]{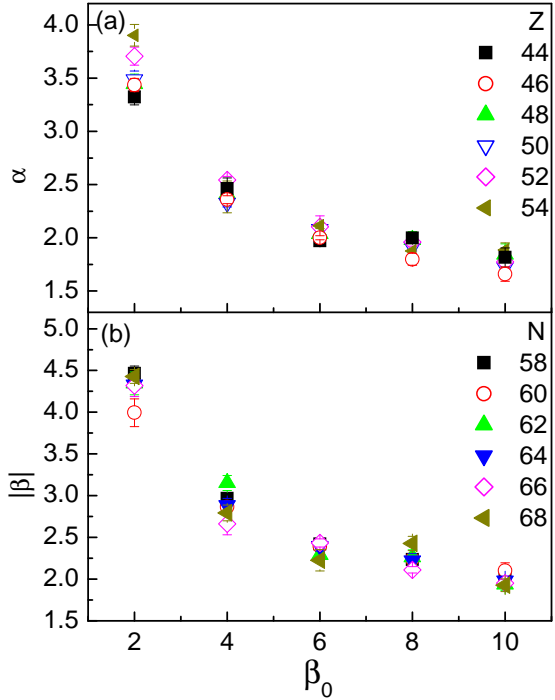}
\vspace{-0.1truein} \caption{\footnotesize (Color online) $\alpha$
and $|\beta|$ as a function of the reduced friction parameter
$\beta_{0}$ for different fission fragments with $Z$ = 44 - 54 and
$N$ = 58 - 68, respectively. The width ($\sigma_{\alpha_0}$) of
the Gaussian probability  is 0.06. } \label{alpha_beta0}
\end{figure}

In addition, the influence of the reduced friction parameter on
the isoscaling parameters is investigated, we use a constant value
of $\beta_{0} = 2, 4, 6, 8$ and 10 instead of one-body dissipation
$\beta_{\rm OBD}$ which was used in above calculations. In
Fig.~\ref{alpha_beta0}(a) and (b), we plot $\alpha$ and $|\beta|$
as a function of $\beta_{0}$ for different elements from $Z$ = 44
to 54 or different isotones from $N$ = 58 to 68, respectively.
Both $\alpha$ and $|\beta|$ decrease with the increasing of the
reduced friction parameter. It shows that $\alpha$ and $\beta$ are
sensitive to the the reduced friction parameter. Larger reduced
friction makes the Brownian particles cost more energies which
will be transferred to the internal energy from ground state to
the scission point than the smaller one, consequently the system
will keep less   memory to the initial entrance channel. In the
viewpoint of the  isoscaling behavior, the isoscaling parameter
shows a decrease with the reduced friction  parameter. Therefore
the study on the isoscaling  behavior to the fission fragment
might be a good tool to explore the friction effect in the fission
dynamics process.

\section{Summary}

In summary, we applied the Langevin  model to investigate the
isoscaling behavior in the dynamical process of compound nuclear
fission. In order to treat the fission fragments, we assume that
the mass asymmetry parameter of the two fission fragments from
the fissioning nucleus is taken from a random number with  a
 Gaussian distribution whose width is
$\sigma_{\alpha_0}$. The simulation illustrates that the isotopic and
isotonic yield
ratios of fission fragments in the dynamical fission channels
 of $^{116}$Sn
+ $^{116}$Sn and $^{112}$Sn + $^{112}$Sn reaction system show the
isoscaling behavior. The terms which are related to the difference
of neutron or proton chemical potential are also extracted. It is
of interesting that the isoscaling parameter $\alpha$ is sensitive
strongly to the Gaussian width $\sigma_{\alpha_0}$ of the mass
asymmetry parameter but $\beta$ looks not. When
$\sigma_{\alpha_0}$ is small, i.e. the fission is almost
symmetric, $\alpha$ increases with the atomic number of fission
fragments, which is similar to the theoretical prediction of a
simple liquid-drop model \cite{Friedman}. In contrary, when
$\sigma_{\alpha_0}$ is large, for instance, $\sigma_{\alpha_0}$ =
0.20, $\alpha$ drops with $Z$ of fission fragments. However, in
the intermediate values of $\sigma_{\alpha_0}$, $\alpha$ shows a
backbending with $Z$ of fission fragments, which is similar to the
observation of the $^{238,233}$U fission data induced by 14 MeV
neutrons \cite{Veselsky2}. In this context, we could say that the
$\alpha$ parameter is sensitive to the asymmetric extent of the
fission-fragments from the fissioning nuclei. However, $\beta$
parameter is insensitive to the width $\sigma_{\alpha_0}$ even
though it always shows the dropping trend with $N$.

In addition, the dependences of beam energy and the reduced
friction parameter for the isoscaling parameters are
systematically investigated. It is found that both $\alpha$ and
$\beta$ drop with beam energy of the projectile as well as the
reduced friction parameter, reflecting the temperature-like
dependence of isoscaling parameters in the fission dynamics. The
disappearance of isospin effect of fission dynamics is expected in
a certain higher beam energy or larger reduced friction parameter.
In general, the isoscaling analysis of the fission data appears to
be a sensitive tool to investigate the fission dynamics.

\section*{Acknowledgement}

 This work was supported in part by the Shanghai Development
 Foundation for Science and Technology under Grant Numbers
 05XD14021 and 03 QA 14066, the  National Natural Science
 Foundation of China  under Grant No 10328259, 10135030 and 10405033, the
 Major State Basic Research Development Program under
  Contract No G200077404.

{}

\end{document}